\documentstyle[11pt]{article}

\begin{document}

\begin{center}

{\bf THE SIGNATURE OF THE NEGATIVE CURVATURE OF THE UNIVERSE IN CMB
MAPS}
\vspace{0.1in}

V.G.GURZADYAN$^{1,2}$ and S.TORRES$^3$
\vspace{0.05in}

1. Sussex University, Brighton, UK; 

2. Yerevan Physics Institute,
Yerevan, Armenia (permanent address);

3. Observatorio Astron\'omico Nacional, Universidad Nacional de
Colombia and Centro Internacional de F\'{\i}sica,
Bogot\'a, Colombia

\end{center}
\vspace{0.05in}


{\bf Abstract} The geodesics followed by cosmic microwave background
(CMB)  photons
show different behaviours depending on the
geometry of space. Namely, the effect of `mixing geodesics'
predicts a distinct signature in CMB maps: threshold-independent
elongated anisotropy
spots in negatively curved geometries. We have found statistically
significant sign for spot elongation in the COBE four year maps.
This can be a direct indication for the negative curvature of the Universe.

\section{The Predicted Feature of CMB Maps}

It appears that the evidence in favor of a negatively curved Universe
is rapidly gaining weight \cite{ColesEllis}.
Here we present
the results of a statistically independent analysis of the four year
COBE-DMR data which favors a
Friedmannian universe with $k=-1$ and hence, $\Omega < 1$ \cite{Guto}.
Our analysis is based on the search of the
signature of the effect of {\it geodesic mixing} in these data.

The study of the projection of geodesics from
(3+1)-dimensional Lorentzian space to a 3D Riemannian one, and
of the behavior of time correlation functions for
geodesic flows on homogeneous isotropic spaces with
negative curvature
leads to the effect of ``geodesic mixing" \cite{GK2,gkbook}
(and references therein). Geodesic flows, being Anosov systems
(locally if the space is not compact),
are exponentially unstable systems
possessing the strongest statistical properties (mixing) and
positive Kolmogorov-Sinai (KS) entropy~$h$.
For a geodesic flow in $k=-1$ Friedmannian Universe
the KS-entropy
depends on the only parameter $a$, the diameter of the Universe:
$h=2/a$ (Lyapunov exponents vanish when $k=0,+1$).

The geodesic mixing has the following observable consequences:
(1) decrease of the amplitude of CMB anisotropy by time,
namely, the anisotropy detected now should be lower than at the
surface of last scattering; (2) flattening of CMB angular autocorrelation
function independent on the initial spectrum at the last scattering
epoch (however, higher anisotropy at measurements with
smaller beam angles is also predicted by mixing);
(3) threshold independent distortion of CMB maps.
Concerning the third effect, the relation between the
quantitative measurement of the distortion  of patterns - spot elongation
parameter,
$\epsilon$, and $\Omega$ is given by \cite{gkbook}
\begin{equation}
\frac{\ln {(1/\epsilon)}}{(\frac{1-\Omega_0}{1+z\Omega_0})^{\frac{1}{2}}} =
\left\{ \begin{array}{ll}
 \alpha/(1-\alpha) [1-(1+z_1)^{1-1/\alpha}] & \alpha<1\\
                                  \ln(1+z_1) &  \alpha=1
 \end{array}
 \right.
\end{equation}
where $\Omega_{0}$ is its present value,
$z_1$ corresponds to the time when matter becomes non relativistic and
$z$ to the decoupling time, $\alpha$ is the power index in the expansion
law of the Universe.
Note that:

(a) The first two effects - the decrease of
anisotropy and the flattening of the autocorrelation function - are
consequences of the negative curvature, while effect (3) on the
map distortion, is a feature only of perturbed $k=-1$ Universe.

(b) Einstein equations do contain information on the curvature
but not on the topology of the Universe; for example, the same $k=0$
curvature can correspond to various topologies - $R^3$, $Tor^3$,
$R^2 \times S^1$, $R^1 \times Tor^2$, etc;

(c) the use of harmonic analysis, being the main tool for study of
CMB properties for $k=0$ models, has intrinsic difficulties
in hyperbolic spaces, due to problems with incompleteness
of the set of eigenfunctions; moreover, their spectrum
can be continuous or even not defined at all depending on the
topology (Sobolev problem).

The use of methods of the theory of dynamical systems
enables one to avoid
in a way the principal difficulty of harmonic analysis.
The followed geodesic mixing as
statistical effect is a result of the photon beam motion
from the last scattering epoch up to the present observer,
independent of the properties of the peaks of
anisotropies at the last scattering surface \cite{Bar}. 
Thus, the effect we are looking brings information on the curvature
but not on the topology of Universe.

\section{Data Analysis: Checking the Prediction}

In order to characterize the presence
of elongated hot spots in CMB maps we have adopted a
strategy that relies on the statistics of topological descriptors,
which has been successfully used
to place important restrictions
on the spectrum of primordial
perturbations $P(k)$ \cite{torres94a, fabbri96},
The analysis presented here is based on the
4 year COBE-DMR maps at 53 GHz \cite{bennett96}.
Signal and noise maps were prepared by
adding and subtracting the two independent DMR channels
(i.e. $0.5A + 0.5B$ and $0.5A - 0.5B$)
and Gaussian smoothing ($\sigma = 2.9^{\circ}$).
The geometric characteristics of hot spots are quite sensitive
to galactic cuts below  $15^{\circ}$ to $20^{\circ}$ but
beyond $20^{\circ}$ our results are stable.
A hot spot with
preset temperature threshold $T_{\nu} = \nu \sigma$ is defined,
where $\sigma$ is the standard deviation of the sky temperature.
The algorithm of the topological analysis of
CMB maps \cite{torres94a,torres95},
the `eccentricity parameter', $\epsilon_{\nu}$,
as the average ratio of the shortest to longest
`axis' of a hot spot for different temperature thresholds.
In order to evaluate the statistical significance of the result
we have performed Monte Carlo studies of noise maps
that take into account instrumental noise
and COBE's beam width;
the same algorithm was used for them (Table 1).
\begin{table}
\caption{Eccentricity parameter of hot spots on COBE maps
($\epsilon_{\nu}^{A+B}$, $\epsilon_{\nu}^{A-B}$)
and comparison with Monte Carlo noise maps
($\epsilon_{\nu}^{MC}$).
$\Delta^{A+B}$ and $\Delta^{A-B}$ denote
the difference (in standard deviations)
between the measured eccentricities
and the mean eccentricity of
noise Monte Carlo maps.}
\vspace{4mm}
\begin{center}
\begin{tabular}{rrrrrrr}
\hline\hline \\
$\nu$ &
$\epsilon_{\nu}^{A+B}$ &
$\epsilon_{\nu}^{A-B}$ &
$\epsilon_{\nu}^{MC}$ &
$\Delta^{A+B}$ &
$\Delta^{A-B}$ \\[2pt]
\hline\rule{0pt}{12pt} \\
1.00 &    0.480 &    0.570 &    0.582  &   3.551 &    0.447 \\
1.25 &    0.547 &    0.590 &    0.617  &   2.271 &    0.886 \\
1.50 &    0.497 &    0.616 &    0.646  &   4.288 &    0.881 \\
1.75 &    0.576 &    0.691 &    0.674  &   2.432 &   -0.429 \\
2.00 &    0.600 &    0.686 &    0.699  &   2.049 &    0.267 \\
2.25 &    0.595 &    0.819 &    0.719  &   2.060 &   -1.675 \\
2.50 &    0.427 &    0.685 &    0.741  &   4.017 &    0.719 \\
2.75 &    0.574 &    0.732 &    0.761  &   1.752 &    0.272 \\
3.00 &    0.507 &    0.705 &    0.776  &   1.815 &    0.480 \\ [2pt]
\hline
\end{tabular}
\end{center}
\end{table}
The $\chi^2$ statistic computed with the 9 data points
in the range $\nu = 1.0 - 3.0$ and the corresponding
noise Monte Carlo points is 5.6 and 73.0 for the $(A-B)$ and
$(A+B)$ maps respectively,
thus indicating the accuracy of
the Monte Carlo simulations.
On the other hand, the high $\chi^2$ obtained
when the data from the signal maps is compared with
noise data is a clear indication of an actual detection
of elongated anisotropy spots.
The average deviation in terms of standard deviations
of $\epsilon_{\nu}^{A+B}$ from
the corresponding Monte Carlo result for the
9 bins considered here is $3\sigma$.

Thus, the excess elongation as possible genuine feature of the
hot spots on COBE maps is well established at least
at threshold levels $\nu$ between 1.0 and 3.0. The
detected signal shows a statistically strong
independence on the threshold, contrary to the case for both
$(A-B)$ and Monte Carlo maps, where $\epsilon_{\nu}$ shows a
clear correlation with threshold.
Fluctuations on the surface of last
scattering might also produce
non-zero ellipticity even in flat universe \cite{bond},
which however will depend on the
threshold, contrary to what
we have observed.
If the effect of elongation detected on CMB maps is due to the
geodesic mixing, as
predicted, then our analysis implies the negative curvature
of the Universe.

The authors are thankful to A.Kocharyan and R.Penrose for valuable
discussions.  V.G. is grateful to J.Barrow and the staff
of Sussex Astronomy Centre for hospitality.
This research is in part funded by COLCIENCIAS of Colombia (2228-05-103-96)
and the Royal Society.
The COBE datasets were developed by the NASA Goddard
Space flight Center under the guidance of the COBE Science
Working Group and were provided by the NSSDC.


\begin{thebibliography}{99}

\bibitem{ColesEllis} Coles, P., Ellis, G. Nature, 370, 609 (1994).

\bibitem{Guto} Gurzadyan V.G. \& Torres, S.  A\&A, 321, 19 (1997).

\bibitem{GK2} Gurzadyan, V.G. \& Kocharyan, A.A.  A\&A, 260, 14
(1992); Int.J.Mod.Phys., D2, 97 (1993).

\bibitem{gkbook} Gurzadyan, V.G. \& Kocharyan, A.A.  {\it Paradigms of the
Large-Scale Universe}, (Gordon and Breach, New York, 1994).

\bibitem{Bar} Barreiro R.B. et al. astro-ph/9612114 (1996).

\bibitem{torres94a} Torres, S.  ApJ, 423, L9 (1994); Ap.Let.Com. 32, 
  95 (1995);

\bibitem{torres95} Torres, S., et al.  MNRAS, 274, 853 (1995).

\bibitem{fabbri96} Fabbri, R. \& Torres, S.  A\&A, 307, 703 (1996)

\bibitem{bennett96} Bennett, C. L., et al.  ApJ, 464, L1 (1996).

\bibitem{bond} Bond, J.R. \& Efstathiou, G.  MNRAS, 226, 655 (1987).

\end{thebibliography}
\end{document}